\documentclass[twocolumn,floats,preprintnumbers,superscriptaddress,nofootinbib,aps]{revtex4-2}

\usepackage{kotex}
\usepackage{graphicx}
\usepackage{bm}
\usepackage{color}
\usepackage{hyperref}
\usepackage[normalem]{ulem}
\usepackage{amsmath}
\usepackage{amssymb}
\usepackage{mathtools}
\usepackage{here}
\usepackage[dvipsnames]{xcolor}

\usepackage{multirow}

\newcommand{\figidx}[1]{(#1)}

\newcommand{\ie}{{{i.e.}}}
\newcommand{\eg}{{{e.g.}}}

\newcommand{\etal}{{\it{et al}.}}
\newcommand{\kT}{k_{B} T}
\newcommand{\enat}{\text{e}}

\def\Eq{Eq.}

\def\Figure{Figure}
\def\Figures{Figures}
\def\Fig{Fig.}
\def\Figs{Figs.}

\extrafloats{100}

\begin{document}

\title{Confinement-Driven Acceleration of First-Passage Rates}

\author{Won Kyu Kim}
\email{wonkyukim@snu.ac.kr}
\affiliation{School of Computational Sciences, Korea Institute for Advanced Study, Seoul 02455, Korea}
\affiliation{Department of Physics and Astronomy, Seoul National University, Seoul 08826, Korea}

\date{\today}

\begin{abstract}
We demonstrate that confinement geometry can act as a rectifier in passive diffusion, optimally accelerating first-passage rates beyond free diffusion. Using analytic theory based on the Fick-Jacobs approach and Brownian dynamics simulations, we find nonmonotonic mean first-passage rates driven by entropy. Through the transmission probability, our findings highlight how confinement optimizes transport dynamics in trap-and-escape processes, with implications for molecular translocation and reaction kinetics in soft matter and biological systems.
\end{abstract}


\maketitle

Kramers’ barrier-crossing rate, characterized by its exponential decay with the energy barrier height, serves as a cornerstone for modeling barrier-crossing processes in stochastic systems~\cite{kramers1940brownian,siegert1951first,hanggi1990reaction,mel1991kramers,van1992stochastic,redner2001guide,kim2015mean,Benichou2014From,metzler2014first,sung2018mean,kim1958mean,szabo_jcp80,park2003reaction,zwanzig1992levinthal,condamin2007first,benichou2010geometry,polizzi2016mean,pal2019first}. 
This approximation leading to the Arrhenius law for chemical reactions assumes large energy barriers, making it less applicable to describing subcellular processes, such as protein folding, where free energy barriers are typically only a few $\kT$ (thermal energy)~\cite{oliveberg2005experimental, schuler2008protein, kim2012weak,wan2024entropy}. In particular, entropic barriers, arising from confinement and restricted molecular configurations, are typically below $10~\kT$, which play a critical role in biomolecular dynamics. 
This interplay enables rapid and dynamic transitions while maintaining selectivity and regulation~\cite{wan2024entropy}---DNA translocation through nanopores involves a loss of configurational degrees of freedom, resulting in an entropic barrier of height $\sim 5~\kT$~\cite{sung1996polymer, bell2016translocation}. The release of neurotransmitters from vesicles of size $40~$nm~\cite{sudhof2004synaptic} through pores of size $10~$nm~\cite{he2007debate} requires overcoming an entropic barrier of height $\sim \kT \ln 4 \approx 1.4~\kT$. 
Ion transport through narrow channels, such as potassium channels, yields entropic barriers of height $3$ to $9~\kT$, arising from dehydration and confinement within selectivity filters~\cite{prost1996shape, wawrzkiewicz2018role, lu2023dehydration}. 
Confinement governs the efficiency and directionality in molecular transport, reaction kinetics, and structural transitions

\begin{figure}[b]
\centering
\includegraphics[width = 0.45\textwidth]{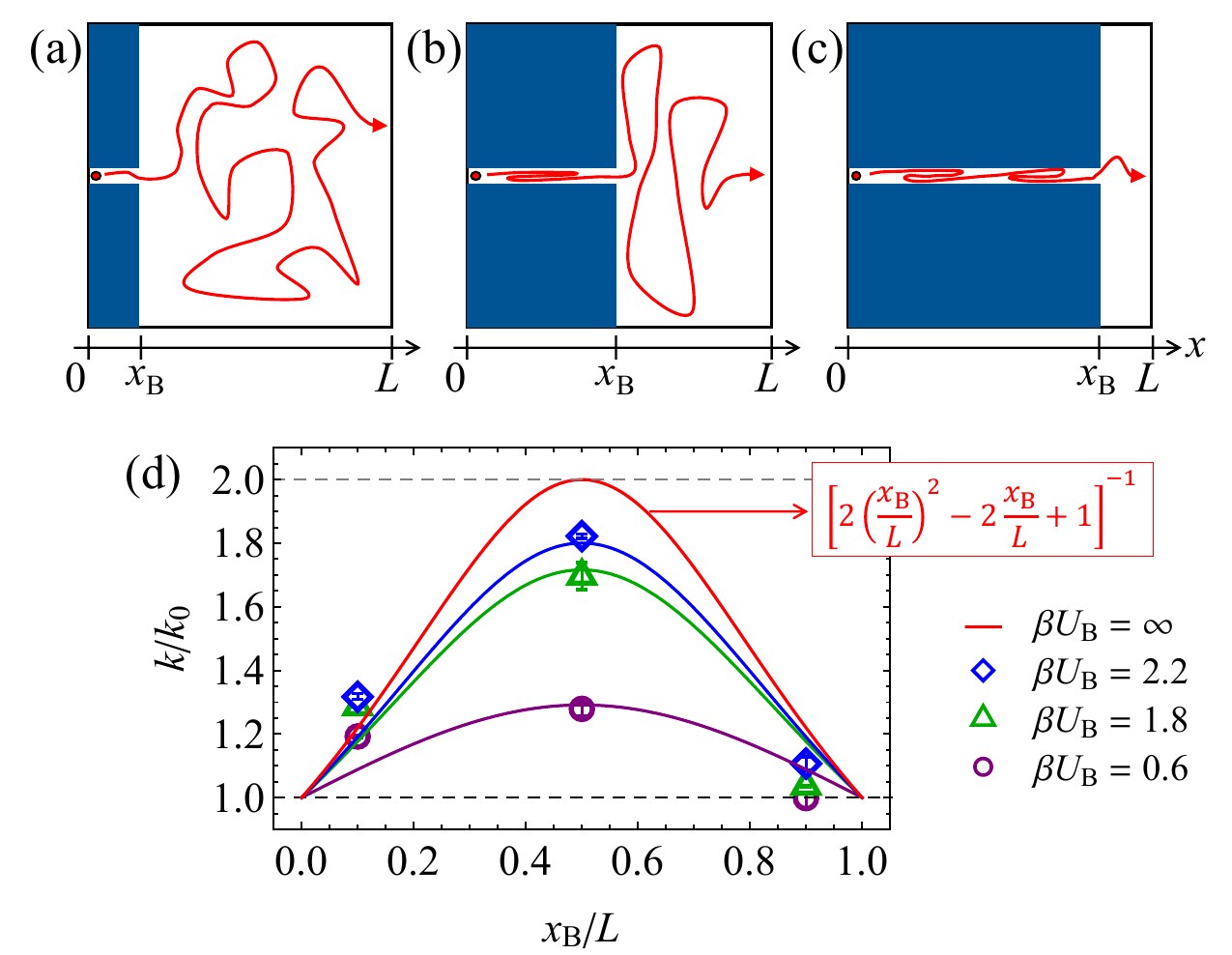}
\caption{Illustration of first passage events through a narrow tunnel of length \figidx{a} $x_{B} \approx 0$, \figidx{b} $x_{B} \approx L/2$, and \figidx{c} $x_{B} \approx L$. \figidx{d} Mean first-passage rate $k(x_{B})/k_0$ for $\beta U_{B} = 0.6$ (circles), $\beta U_{B} = 1.8$ (triangles), and $\beta U_{B} = 2.2$ (diamonds), obtained from 2D Brownian dynamics simulations with a tunnel width $w = L\enat^{-\beta U_{B}}$. The rate $k_0$ corresponds to the tunnel-free diffusion rate. Solid lines represent the analytic result by the Fick–Jacobs approximation. Top red line depicts $k/k_0 = 1/[2(x_{B}/L)^2 - 2x_{B}/L + 1]$ in the limiting case for $U_{B} \to \infty$. 
}\label{fig5}
\end{figure}

\begin{figure*}
\centering
\includegraphics[width = 0.77\textwidth]{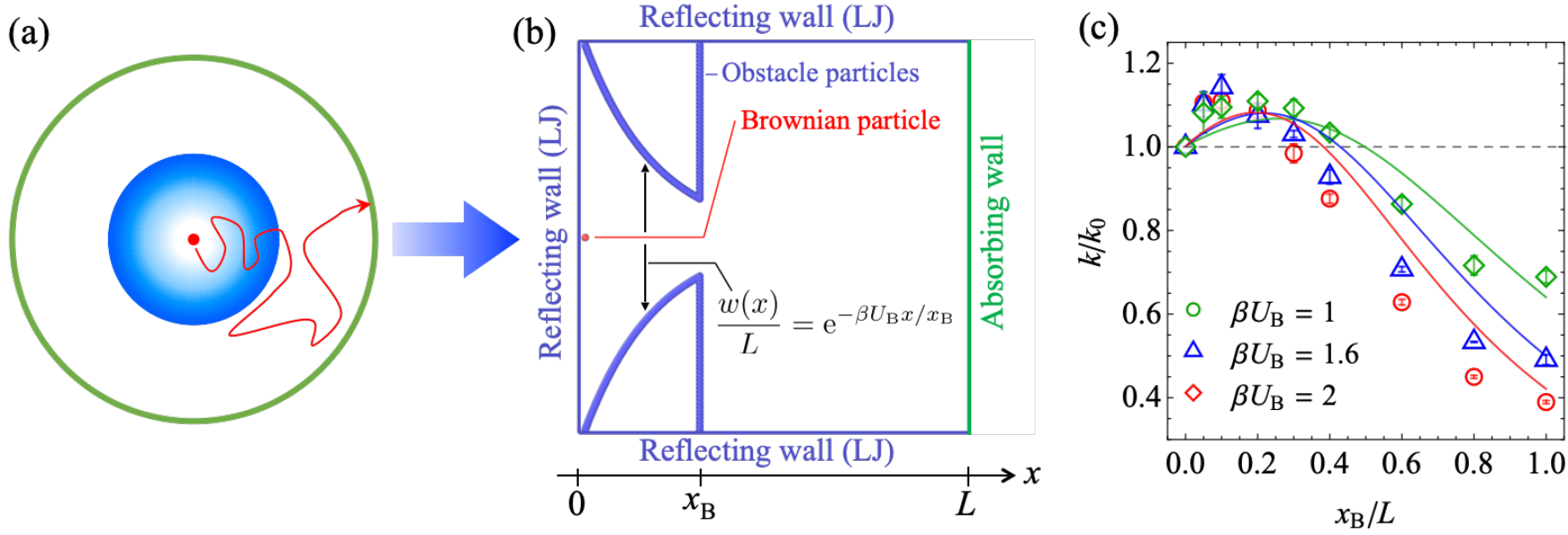}
\caption{
\figidx{a} Example of trap-and-escape process, featuring a trapping vesicle composed of radially inhomogeneous confinement. \figidx{b} 2D Brownian dynamics simulations with trapping confinement formed by immobile hard particles, where the accessible cross-section length is $w(x) = L \exp(-\beta U_{B} x / x_{B})$.
\figidx{c} Symbols: Normalized mean first-passage rate $k/k_0$ as a function of $x_{B}$ (barrier position) and $U_{B}$ (barrier height at $x_{B}$) obtained from the simulations, displaying a nonmonotonic dependence and maximization of $k$ beyond the trap-free case $k_0$. Error bars indicate the standard error calculated from three independent datasets. 
Solid lines depict $k/k_0$ numerically calculated using the modified Fick-Jacobs approximation~\cite{reguera2001kinetic}.
}\label{fig2_}
\end{figure*}

In this work, we explore stochastic escape processes over moderate entropic barriers arising from confinement. In particular, we investigate how confinement can enhance escape dynamics in terms of the mean first-passage rate (MFPR), with a target boundary located in outer space [\eg, \Figs~\ref{fig5}\figidx{a}-\figidx{c} and \ref{fig2_}\figidx{a}]. This process can be aptly termed \textit{trap-and-escape}, a phenomenon commonly observed in biology and soft matter~\cite{schuss2007narrow, elias2007synaptic,zhou2008macromolecular}. Examples include molecular diffusion in cellular microdomains (e.g., ribosome exit tunnels~\cite{yu2023geometric}), neurotransmitter transport between synaptic membranes, and molecular transport across porous membranes with selective permeability~\cite{shaw2007geometry,roa2017catalyzed,kim2019prl,kanduvc2020modeling,kim2020tuning,kim2022permeability}, reflecting the relevance of this study in understanding the role of thermal activation and confinement in stochastic transport processes.

{\it Confinement-driven rectification.}---Let us first explore an idealized two-dimensional (2D) example that illustrates how confinement can enhance the MFPR. Consider a narrow tunnel of length $x_{B}$ positioned on the left side of a box of length $L$, as illustrated in \Figs~\ref{fig5}(a)-(c). Starting at $(x=0, y=L/2)$, a point-like Brownian particle escapes the tunnel by passing through $x=x_{B}$ and subsequently reaches a target boundary at $x=L$, while all the other boundaries are reflective. In the absence of the tunnel (i.e., only a box), the corresponding mean first-passage time is $\tau_0 = L^2 / (2D)=1/k_0$~\cite{berezhkovskii2007diffusion}, where $D$ is the diffusivity. In the presence of a tunnel whose width is sufficiently narrow, the probability of an escaped particle reentering the tunnel becomes negligible. In this limit, the mean first-passage time decomposes into two contributions: (i) the time spent in the tunnel, $\tau_t = x_{B}^2 / (2D)$, and (ii) the time spent outside the tunnel, $\tau_f = (L - x_{B})^2 / (2D)$. Thus, the total mean first-passage time is $\tau = \tau_t + \tau_f$, and the corresponding MFPR is $k / k_0 = 1 / [2(x_{B}/L)^2 - 2x_{B}/L + 1]$. This Lorentzian-like function attains its maximum at the location parameter $x_{B}=L/2$, at which it reaches a maximum, as shown by the red line in \Fig\ref{fig5}(d). The latter is {\textit{twice}} as fast as the tunnel-free rate $k_0$, exemplifying the confinement-driven rectification and acceleration of the MFPR.

To describe effects of finite tunnel width on the MFPR, we apply the Fick–Jacobs (FJ) approach~\cite{zwanzig1992diffusion,reguera2001kinetic,berezhkovskii2007diffusion,pompa2022first}, which approximates confinement with rotational symmetry as an entropic free energy in 1D. For an accessible tunnel width at position $x$ denoted by $w(x)$, the entropic free energy in units of $\kT=1/\beta$ is  $\beta U(x) = -\ln [w(x)/L]$~\cite{zwanzig1992diffusion}. 
The corresponding MFPR is then given by $k_{FJ}/k_0 = 1/\int_{0}^{L}\text{d}x\int_{0}^{x}\text{d}x'\, [ 1+w'(x)^2/4 ] ^{1/3}w(x')/w(x)$~\cite{jackson1963effective,szabo_jcp80,kim2015mean}, where $w'(x)=\text{d}w(x)/\text{d}t$.
Considering a constant tunnel width $w = L\enat^{-\beta U_{B}\theta(x_{B}-x)}$, where $\theta(x)$ is the Heaviside step function, the FJ approach yields $k_{FJ}/k_0 = 1/[2(x_{B}/L)^2 - 2x_{B}/L + 1 + 2\enat^{-\beta U_{B}}(1 - x_{B}/L)x_{B}/L]$ (see Supplemental Material~\cite{sm} for detailed derivation). This expression recovers the previous $k$ in the limit $U_{B}\rightarrow \infty$, implying that the last term in $k_{FJ}/k_0$ is responsible for the reentrance process. The purple ($\beta U_{B} = 0.6$), green ($\beta U_{B} = 1.8$), and blue ($\beta U_{B} = 2.2$) lines in \Fig~\ref{fig5}(d) depict the $k_{FJ}(x_{B},U_{B})/k_0$, which show good agreement, particularly for $x_{B}=0.5L$, with results from 2D Brownian dynamics (BD) simulations involving a tunnel and boundaries formed by immobile hard particles. 
For hard-particle interactions between all particles
, we use the Weeks–Chandler–Andersen (WCA) potential~\cite{wca} (Lennard–Jones potential truncated at its minimum and shifted to zero). 
Additional simulation details are provided in the Supplemental Material~\cite{sm}. 


The results demonstrate that confinement can optimally suppress reentrance processes, acting as a {\textit{rectifier}} in passive diffusion, thereby enhancing the MFPR.
While the MFPR predicted by the FJ approximation is symmetric around $x_{B} = 0.5L$, the particle-based simulation results exhibit slight asymmetry, which will be discussed further in the following.

{\it Enhancement of MFPR by position-dependent confinement.}---To describe trap-and-escape processes more generally, now we consider a position-dependent $w(x)$.
An example for this may include a trapping capsule composed of radially inhomogeneous confinement, illustrated in \Fig~\ref{fig2_}\figidx{a}.
As a tractable, simplified analogous version, we study the latter via a 2D system shown in \Fig~\ref{fig2_}\figidx{b}.
We perform BD simulations, similarly to the previous section, using immobile hard particles in a box of length $L$. 
The hard particles, shown in blue in \Fig~\ref{fig2_}\figidx{b}, overlap to form trumpet-shaped boundaries along $y=L(1\pm \enat^{-\beta U_{B} x/x_{B}})/2$
in $0 \leq x \leq x_{B}$, creating an exponentially decreasing tunnel of width $w(x) = L \exp(-\beta U_{B} x/x_{B})$. At $x=x_{B}$, the obstacles form vertical walls, determining the size of a tunnel exit $w(x_{B}) = L \exp(-\beta U_{B})$, \ie, a trapping barrier.  
We compute the MFPR from $(x=0,y=L/2)$ to $(x=L,\forall y)$ for varying barrier positions $x_{B}$ and heights $U_{B}$. The reference MFPR, $k_0$, is also computed from simulations without a barrier (box only), of which the mean first-passage time corresponds to $\sim 10^7$ time steps in our simulation setup. Additional simulation details are provided in the Supplemental Material~\cite{sm}.

An intriguing MFPR emerges: Symbols in \Fig~\ref{fig2_}\figidx{c} show the obtained $k/k_0$ as a function of $x_{B}$ at fixed values of $\beta U_{B} = 1, 1.6$, and $2$. We observe that $k$ reaches a maximum near $x_{B} = 0.2L$, enhancing the MFPR by approximately $20\%$ compared to the barrier-free diffusion case. The MFPR is nonmonotonic and decreases below unity as $x_{B}$ further increases.
The corresponding crossover position of $k/k_0$ depends on $\beta U_{B}$, indicating that as $\beta U_{B}$ increases (smaller exit size), the peak position of $k(x_{B})$ tends to approach $x_{B} = 0$. 

The enhancement of MFPR due to position-dependent confinement is nonmonotonic and considerably asymmetric, compared to the previous tunnel case.

\begin{figure}
\centering
\includegraphics[width = 0.32\textwidth]{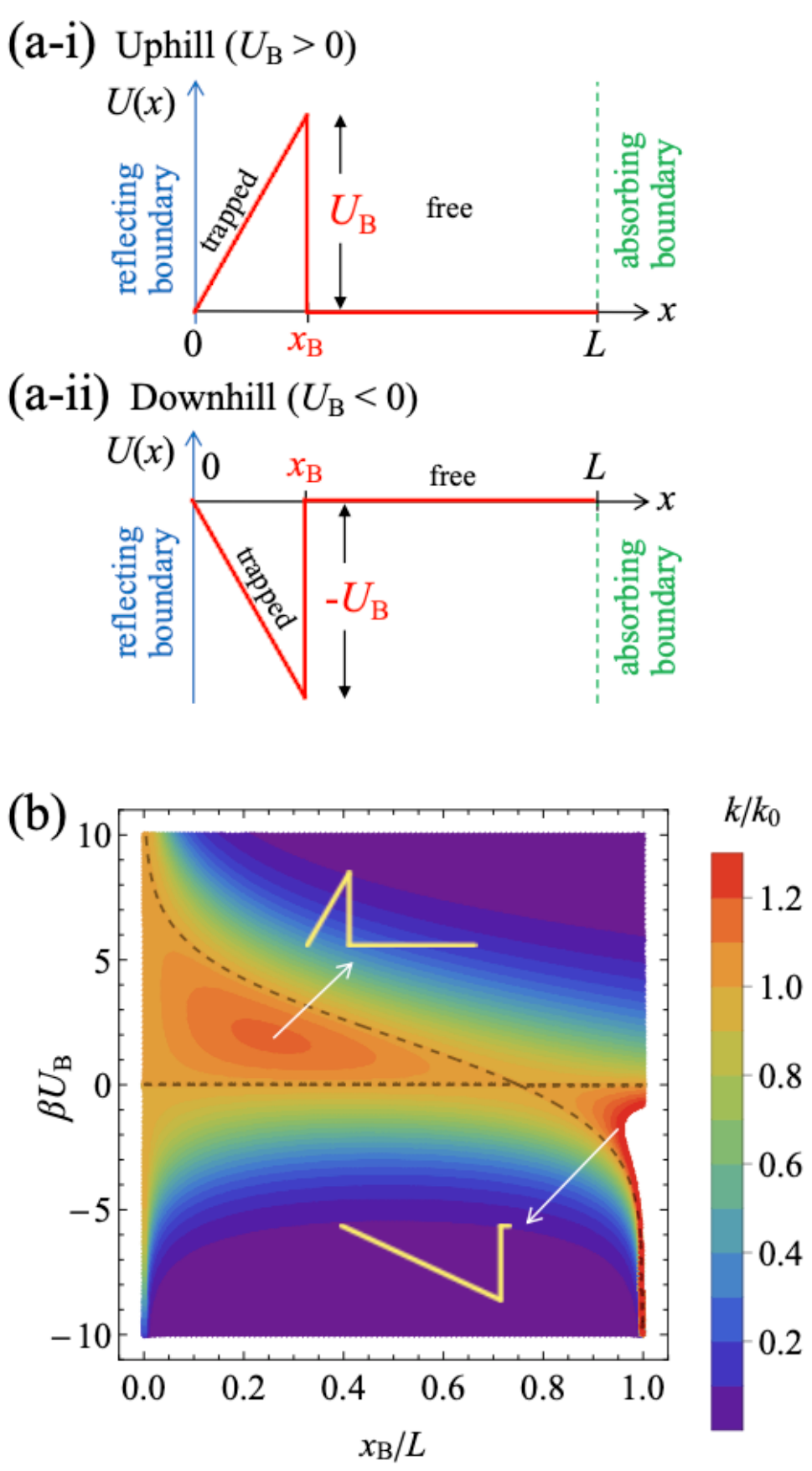}
\caption{
\figidx{a} Free energy landscape $U(x) = \theta(x_{B}-x) U_{B} x / x_{B}$, where $\theta(x)$ is the Heaviside step function. For $U(x_{B})>0$, the barrier is uphill (a-i), while for $U(x_{B})<0$, the well is downhill (a-ii).
\figidx{b} Contour plot of mean first-passage (from $x=0$ to $x=L$) rate $k(x_{B}, U_{B})/k_0$ derived in \Eq~\eqref{eq1}. The dashed lines represent contour lines at $k/k_0=1$, above which the mean first-passage rate is faster than that in free diffusion (orange--red region), such as at the peak near $x_{B} \approx 0.3L$ and $U_{B} \approx 2\kT$. The arrows indicate representative free energy shapes at two distinct $(x_{B}, U_{B})$, each yielding rapid first passages.
}\label{fig3_}
\end{figure}

{\it Mapping to 1D barrier-crossing processes.}---To understand the observed confinement-induced enhancement of MFPR, we apply the FJ approach. This converts the trumpet-shaped confinement into a linear function $U(x) = \theta(x_{B}-x) U_{B} x / x_{B}$, as shown in \Fig~\ref{fig3_}\figidx{a}. 
For $U_{B} > 0$, the shape of the trapping {\it{barrier}} is uphill [\Fig~\ref{fig3_}(a-i)], while for $U_{B} < 0$, the trapping {\it{well}} is downhill [\Fig~\ref{fig3_}(a-ii)].

Note that, since $w(x)$ is position dependent, here we consider the modified FJ approach~\cite{reguera2001kinetic}. The corresponding mean first-passage time is $\tau = \int_{0}^{L}\text{d}x\int_{0}^{x}\text{d}x'\,\enat^{-\beta U(x')}\enat^{\beta U(x)}/D(x)$, which includes a correction factor in $D(x) \approx D/[1 + (\text{d}w(x)/\text{d}x)^2/4]^{1/3}$~\cite{zwanzig1992diffusion,reguera2001kinetic}, which is crucial when $w(x)$ changes considerably. Additionally, depending on the rotational asymmetry of confinement, further corrections may arise~\cite{pompa2022first}. However, it is known that for wide-to-narrow confinement that is rotationally symmetric, as considered in this work, the above second-order correction provides a good approximation~\cite{berezhkovskii2007diffusion,pompa2022first}. 
We calculate $k/k_0$ including the correction in $D$ consistently, shown by the solid lines in \Fig~\ref{fig2_}\figidx{c} (for detailed derivation, see the Supplemental Material~\cite{sm}). 
\Figure~\ref{fig2_}\figidx{c} compares the results for the MFPR obtained from 2D confinement simulations with those from the modified FJ approximation, showing good qualitative agreement, particularly in capturing the nonmonotonic behavior.
The discrepancy between the two arises from the finite particle size in the simulations and the lack of higher-order correction terms in the theory.

\begin{figure*}
\centering
\includegraphics[width = 0.83\textwidth]{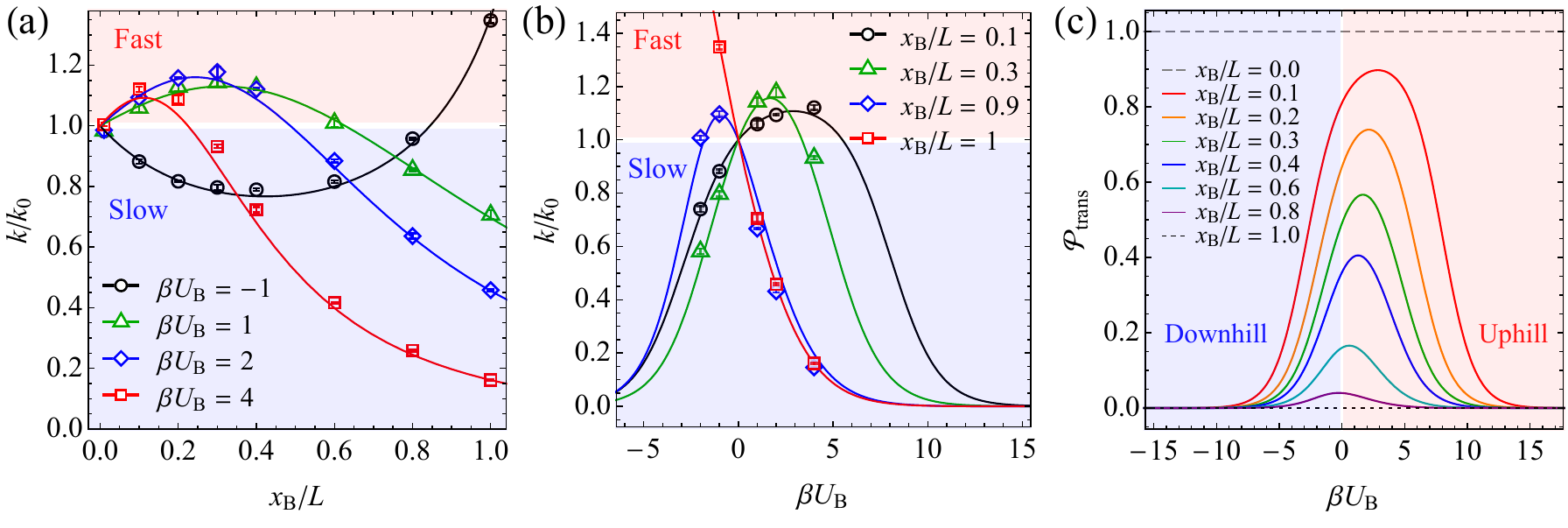}
\caption{
Mean first-passage rate $k/k_0$ as a function of \figidx{a} barrier position $x_{B}$ and \figidx{b} barrier height $U_{B}$. The solid lines represent the theoretical prediction in \Eq~\eqref{eq1}, while symbols show the rate $k/k_0$ obtained from one-dimensional Brownian dynamics simulation data. The red shaded region indicates a fast rate $(k/k_0 > 1)$, while the blue shaded region represents a slow rate $(k/k_0 < 1)$, in comparison to the free-diffusion rate $(k/k_0 = 1)$. Error bars indicate the standard error calculated from three independent datasets.
\figidx{c} Transmission probability $\mathcal{P}_\text{trans}$ for occupying the free region $x_{B} \leq x \leq L$. A positive, finite trapping energy barrier maximizes the transmission probability for $0 < x_{B}/L < 1$.
}\label{fig4_}
\end{figure*}
 
Although the modified FJ approximation can yield an analytic expression after a lengthy derivation, we proceed by assuming a constant diffusivity $D$ (\ie, the original FJ approach) to better understand the nonmonotonic nature of $k$. This simplification does not significantly alter the main result and helps to physically rationalize the observed nonmonotonic behavior. It also provides a representative concise 1D barrier-crossing model for trap-and-escape processes. With the simple linear form of $U(x)$ and constant $D$, we derive an analytic expression for the normalized MFPR, which takes the form of a Lorentzian-like function:
\begin{eqnarray}\label{eq1}
\frac{k}{k_0} = \frac{\mathcal{A}}{\left(\frac{x_{B}}{L} - \lambda^\ast \right)^2 +  \mathcal{A} - {\lambda^\ast}^2},
\end{eqnarray}  
where the location parameter $\lambda^\ast$ is
\begin{eqnarray}\label{eq2}
\lambda^\ast  &=& \mathcal{A} (\mathcal{H}+1),
\end{eqnarray}  
with
\begin{eqnarray}
\mathcal{A} &=& \frac{\beta U_{B}}{2  \mathcal{H} (\beta U_{B}-\enat^{\beta U_{B}})+\beta U_{B}-2}, \nonumber \\
\mathcal{H} &=& \frac{\enat^{-\beta U_{B}}-1}{\beta U_{B}}. \nonumber
\end{eqnarray}  
For a detailed derivation and the charateristics of the location parameter $\lambda^\ast (U_{B})$, see the Supplemental Material~\cite{sm}.

The strong nonmonotonic dependency of $k/k_0$ on $x_\text{B}$ and $U_\text{B}$ is highlighted by the contour plot in \Fig~\ref{fig3_}\figidx{b}. The landscape of $k/k_0$ is notably varied, with a peak of $k/k_0 \approx 1.2$ at $x_\text{B} \approx 0.3L$ and $U_\text{B} \approx 2\kT$. This behavior is in line with the results found from the 2D confinement simulations [\Fig~\ref{fig2_}(c)]. Dashed lines represent contour levels at $k/k_0 = 1$, above which (orange--red regions) passages are faster than in barrier-free diffusion. Arrows in \Fig~\ref{fig3_}\figidx{b} indicate two distinct free energy barrier shapes at different $(x_\text{B}, U_\text{B})$, each yielding fast MFPRs. We observe that an uphill barrier with a finite position and height $(U_\text{B} > 0~\wedge~x_\text{B} > 0)$ can maximize the MFPR, while a downhill well with a finite position and height $(U_\text{B} < 0\wedge~x_\text{B} \approx L)$ can also enhance the MFPR.

We validate our analytic result in \Eq~\eqref{eq1} via additional 1D Brownian dynamics simulations using the same linear $U(x)$ (for more details, see the Supplemental Material~\cite{sm}).

\Figures~\ref{fig4_}(a) and (b) show $k(x_{B})/k_0$ and $k(U_{B})/k_0$, respectively, where solid lines denote \Eq~\eqref{eq1} and symbols represent the 1D simulation results, confirming the nonmonotonic feature of MFPR with respect to both $x_{B}$ and $U_{B}$. This extends our findings from the 2D confinement simulations, particularly in its dependence on $U_{B}$.
The MFPR as a function of $U_{B}$, shown in \Fig~\ref{fig4_}(b), contrasts sharply with conventional barrier-crossing behavior. For large $U_{B}$, the rate $k$ converges to the Kramers rate, characterized by exponential decay as depicted by the red line (rectangle symbols) for $x_B=L$. However, at finite $U_{B}$ and $x_B$, the rate can be maximized beyond the barrier-free diffusion rate ($k/k_0 >1$).

Therefore, the theoretical result in \Eq~\eqref{eq1} supports the observations in \Fig~\ref{fig2_}(c). 
To highlight this, we calculate the transmission probability for escape, defined as the probability of being in the barrier-free region $x_{B} < x \leq L$, which leads to
\begin{eqnarray}\label{eq:3}
\mathcal{P}_\text{trans}
&=& \frac{k}{k_0}\left( 1 - \frac{x_B}{L}\right)^2.
\end{eqnarray} 
This is derived from a solution of the Smoluchowski equation (for detailed derivation, see the Supplemental Material~\cite{sm}).
Note that \Eq~\eqref{eq:3} can be generally applied to trap-and-escape processes, provided $k$ is known.

\Figure~\ref{fig4_}(c) shows the $\mathcal{P}_\text{trans}(U_{B})$ at several fixed values of $x_{B}$.
The long-dashed line represents the case when $x_{B} = 0$ (the barrier-free case), resulting in $\mathcal{P}_\text{trans} = 1$.
Conversely, the short-dashed line corresponds to the fully trapped case with $x_{B} = L$, yielding $\mathcal{P}_\text{trans} = 0$.
The highly nonmonotonic behavior of $\mathcal{P}_\text{trans}$ emerges only at finite, intermediate values of $x_{B}$, where the presence of a finite barrier ($U_{B} > 0$) maximizes $\mathcal{P}_\text{trans}$. Notably, the maximum $\mathcal{P}_\text{trans}$ exceeds $0.5$ within the range $0 \leq x_{B} \lessapprox 0.3$, emphasizing the rectification mechanism induced by finite trapping (uphill barrier) at $U_{B} \approx 2\kT$ and $x_{B} \approx 0.3L$.
The nonmonotonic nature of $\mathcal{P}_\text{trans}$ with respect to $U_{B}$ is entirely governed by $k(U_{B})$, as shown in \Eq~\eqref{eq:3}.

Our findings demonstrate confinement-induced rapid first-passage events, functioning as a geometry-dependent rectifier~\cite{shaw2007geometry,lee2021geometry}, which can emerge in trap-and-escape processes. This phenomenon is also closely related to transport across porous membranes with selective permeability~\cite{shaw2007geometry,kim2019prl,kim2020tuning,kim2022permeability}, which is ubiquitous in various biological and physical processes. Previously, R.~S.~Shaw \etal~\cite{shaw2007geometry} observed that a thin membrane with asymmetric pores, positioned in the middle of a box containing small particles, results in more particles residing on the side with the smaller pore ends within the observation time (see Fig.~5 in their work). This behavior was attributed to dissipative effects arising from particle-wall collisions.
In addition, previous studies have reported a similar enhancement of MFPR in highly asymmetric one-dimensional bistable systems~\cite{palyulin2012finite, palyulin2013speeding, chupeau2020optimizing}, which describe the kinetics of reaction paths between two energy minima. The shape of the asymmetric bistable potential plays a critical role in determining the fast MFPR, in which the driving force (the slope of the potential) after barrier crossing is crucial. In this work, we identify a distinct type of confinement-induced rectification and optimization of first-passage rates within the framework of simulations with actual confinement and the Fick-Jacobs approach, where the dynamics are driven by thermal activation and entropy, without any dependency on the potential energy after escape.

{\it Conlusion.}---We have investigated confinement-driven acceleration of mean first-passage rates in trap-and-escape events. Brownian dynamics simulations with particle-based confinement reveal that even in passive systems, mean first-passage rates exceeding those of free diffusion can be achieved through optimally designed confinement.
While fast mean first-passage rates have previously been reported in bistable systems under gravitation~\cite{palyulin2012finite} and in highly asymmetric potentials~\cite{chupeau2020optimizing}, our work focuses on the critical role of confinement geometry in escape kinetics, which turns out to function as a rectifier. 
Using analytical results based on the Fick--Jacobs approach, we rationalized this geometry-induced rectification mechanism, which exhibits a nonmonotonic dependence on barrier position and height, consistent with our simulation results. The main mechanism underlying the latter is presented through the transmission probability.
Our findings demonstrate that rectification effects, typically associated with active, nonequilibrium systems, can also emerge in steady-state systems through thermal activation and confinement. 
These findings have important implications for molecular transport and reaction dynamics, as the observed optimal barrier heights are on the order of a few $k_BT$, a scale pertinent to biological processes such as biomolecular translocation and protein folding~\cite{sung1996polymer,oliveberg2005experimental, schuler2008protein, kim2012weak}. 
Future work could investigate confinement effects in more complex systems, such as biopolymer dynamics or densely packed molecules, thereby extending the applicability of our findings to soft matter and biological systems.

I thank Roland R. Netz, Jaeoh Shin, and Changbong Hyeon for fruitful discussions. I also acknowledge the support from the KIAS Individual Grants (CG076002) at the Korea Institute for Advanced Study (KIAS), as well as the Center for Advanced Computation at KIAS for providing computing resources for this work.


%


\end{document}